\documentclass[12pt]{article}

\usepackage{epsf}
\usepackage[dvips]{graphicx}
\usepackage{amsmath,amssymb}
\usepackage{floatflt,wrapfig}

\begin{document}

\begin{center}
{\bfseries MASS HIERARCHY OF COLLISIONAL ENERGY LOSS}

\vskip 5mm

R.S. Kolevatov$^{\dag}$\footnote{On work in collaboration with Urs Wiedemann, CERN, Theoretical division, [arXiv:0812.0270].}

\vskip 5mm

{\small {\it
St-Petersburg State University
}
\\
$\dag$ {\it
E-mail: rodion@mail.cern.ch
}}
\end{center}

\vskip 5mm

\begin{center}
\begin{minipage}{150mm}
\centerline{\bf Abstract}
Collisional parton energy loss is revisited within a simple model assuming incoherent elastic scattering of on-shell projectile partons on partonic constituents of the QGP with HTL screening. The thermal motion of plasma particles is carefully taken into account. Results on $dE/dx$ are found to be consistent with other authors. There is a significant discrepancy in the energy loss pattern for the cases with thermal motion on and off, which illustrates the importance of taking the kinematics into account exactly. The dependence on the mass of the partons forming the plasma is included in the calculations and its influence on the collisional energy loss is studied. The mass hierarchy of collisional energy loss is found to have a strong dependence on the mass introduced for plasma particles. Due to difference in the mass hierarchy with radiative energy loss, the collisional one when included increases the relative suppression of heavy quarks compared to light quarks.
\end{minipage}
\end{center}

\vskip 10mm

\section{Introduction}

'Hard probes' that is observables connected with high-$p_t$ particles, produced at central rapidity in a collision of heavy ions, are regarded as an effective instrument for testing properties of the matter created in that collision. The production process of a high-$p_t$ parton has a very short typical time scale, of the order of $1/p_t$, while the typical hadronization time scale is large, which means that partons hadronize outside of the medium created. Hence, provided one knows the mechanisms of high-$p_t$ parton production and hadronization, information about the parton propagation between production and hadronization can be used to extract certain properties of the medium. 

One of the most striking features observed in the collision of heavy ions at RHIC is a strong suppression in high $p_t$ inclusive cross sections for light hadrons and non-photonic electrons, so-called 'jet quenching', which is commonly attributed to energy loss of partons (quarks and gluons) propagating in medium before the hadronization. Generally speaking there are two kinds of processes which lead to the energy loss of fast partons, namely elastic and inelastic. Inelastic interactions accompanied with gluon radiation due to kinematic reasons are the dominant mechanism at high parton momentum they are found to be capable to describe observed supression in pionic cross sections, and until not too long ago the contribution of elastic processes was assumed to be negligible. Recently, RHIC data on non-photonic electrons have raised the question whether a radiative mechanism alone is sufficient to explain the energy loss for heavy quarks (especially $b$). This has revived the interest in elastic collisions as an additional source of parton energy loss.

From the medium side, radiative energy loss is fully characterized by a single parameter $\hat q$, which is the average transverse momentum squared transferred to the parton by the medium per unit path length. The relative amount of the energy loss of massive quarks is fully determined by the ratio 
$m/E$ of the projectile. This fact sets up mass hierarchy of the radiative energy loss: the energy loss is smaller for larger projectile mass, and this mass hierarchy stays unchanged with the variation of $\hat q$. \looseness=-2

In the existing models of collisional energy loss (e.g. \cite{Djordjevic:2006tw,Wicks:2005gt}) the medium is usually described as a thermalized bath of quarks and gluons. Here energy loss is fully characterized by the temperature of the medium which determines its density and thus the rate and strengths of interactions. However it is still unclear from the experimental data whether this picture is fully applicable in heavy ion collisions.

At the same time, in contrast with radiative, collisional energy loss and its mass hierarchy in particular, should be sensitive not only to the density of the medium, but also to the ability of the medium to absorb recoil. One can expect this from simple kinematic reasons. Consider a classical example of a collision of two billiard balls with different masses, the projectile ball moving and the target one being at rest. In a case of light projectile and heavy target, the projectile will be simply reflected at some angle without any considerable change in the absolute value of its momentum. And vice versa, a heavy  projectile just will not feel the light target and will simply roll over it. However, when masses of the projectile and the target are equal, we could have a 100\% projectile energy loss in a head-on collision, at other centralities energy loss will still be considerable. At the same time, for a relativistic projectile with $m/E \ll 1$ the dependence of energy transfer in a collision  on the ratio of the masses should disappear. It is important to stress that the arguments above follow from kinematic considerations only. Thus one can expect a non-trivial dependence of the collisonal energy loss of a fast projectile propagating through the medium on the particular properties of the latter. \looseness=-2

Here we address this question is some detail. We model the ability of the medium to absorb recoil simply by attributing its partons some mass $m_t$ which we consider as a parameter.

\section{Model of collisional energy loss}

In recent models proposed to describe collisional energy loss, the medium is modelled either by quarks and gluons in thermal equilibrium \cite{Djordjevic:2006tw,Wicks:2005gt}, or, to simplify consideration, by a number of massive scatterers at rest \cite{Wicks:2007zz}. In the spirit of these approaches we model the medium as a set of quarks and gluons with uniform distribution in space and isotropic distribution of momentum. These quarks and gluons are given a mass $m_t$ which we consider as a model parameter. We additionally assume that the projectile $p$ scatters incoherently off partons which constitute the medium. Then the amount of energy transferred by the projectile $Q$ to the medium is proportional to the number of scattering events, that is
\begin{equation}
\frac{d\Delta E_Q}{dx}= \frac{1}{dx/dt} \frac{dN}{dt} \langle \Delta E\rangle_{\text{single scattering}}
\end{equation}
which, in turn, is written in terms of the elastic cross section integrated over the momentum directions of the incoming particle and the momentum distributions of quarks and gluons ($n_q(k)$, $n_g(k)$) in the medium:
\begin{equation}
\frac{d\Delta E_Q}{dx}=\frac{1}{v_Q} \int dp_f (E-E_f(p_f)) \int k^2 dk \left(n_q(k) + \frac{9}{4} n_g(k)\right) 
\frac{d\sigma^{\rm int}_{Qq}(k,p_f)}{dp_f}.
\end{equation}
In the elastic $Q-q$ scattering cross section, we take a leading $t$-channel exchange with hard thermal loop regularization for the matrix element \cite{Kalashnikov:1979cy,Klimov:1982bv}. This is the starting point for a number of calculations \cite{Thoma:1990fm,Braaten:1991we}. We approximate the quark-gluon cross section as
$\frac{d\sigma^{\rm int}_{Qg}( k,p_f)}{dp_f} =
\frac{C_A}{C_F}\frac{d\sigma^{\rm int}_{Qq}( k ,p_f)}{dp_f}$. The mass $m_t$ of the scatterers in the medium is a free parameter, which we vary. We do not make any assumption on the smallness of masses both in matrix element and phase space volume in computing cross sections.

For the momentum distributions of quarks and gluons in the medium we take a thermal one 
\begin{equation}
 n_q(k)=\frac{1}{(2\pi)^3}\frac{12 n_f}{e^{k/T}+1}; \quad n_g(k)=\frac{1}{(2\pi^3)}\frac{16}{e^{k/T}-1}\label{distr}
\end{equation}
with $n_f=2$. We keep distribution the same irrespective of $m_t$ in order to disentangle effects originating due to recoil properties of the medium from other effects which appear due to changes in $m_t$ (decrease of densities, modification of momentum distributions, etc.).

As partonic projectiles, we consider
light quarks ($m_q=200$~MeV), charm quarks ($m_c=1200$~MeV) and bottom quarks 
(${m_b=} 4750$~MeV). We make calculations for the following set of parameters: The temperature, which enters the matrix element through the propagator and which fixes the momentum distributions $n_q(k)$ and $n_g(k)$ is taken to be $T=225$~MeV. The in-medium path length of the projectile is $L=5$~fm. We make calculations for $m_t=200,$ 450, 680 and 1000~MeV.

\begin{figure}[!b] 
 \noindent \parbox[t]{0.47\hsize}{\includegraphics[width=\hsize]{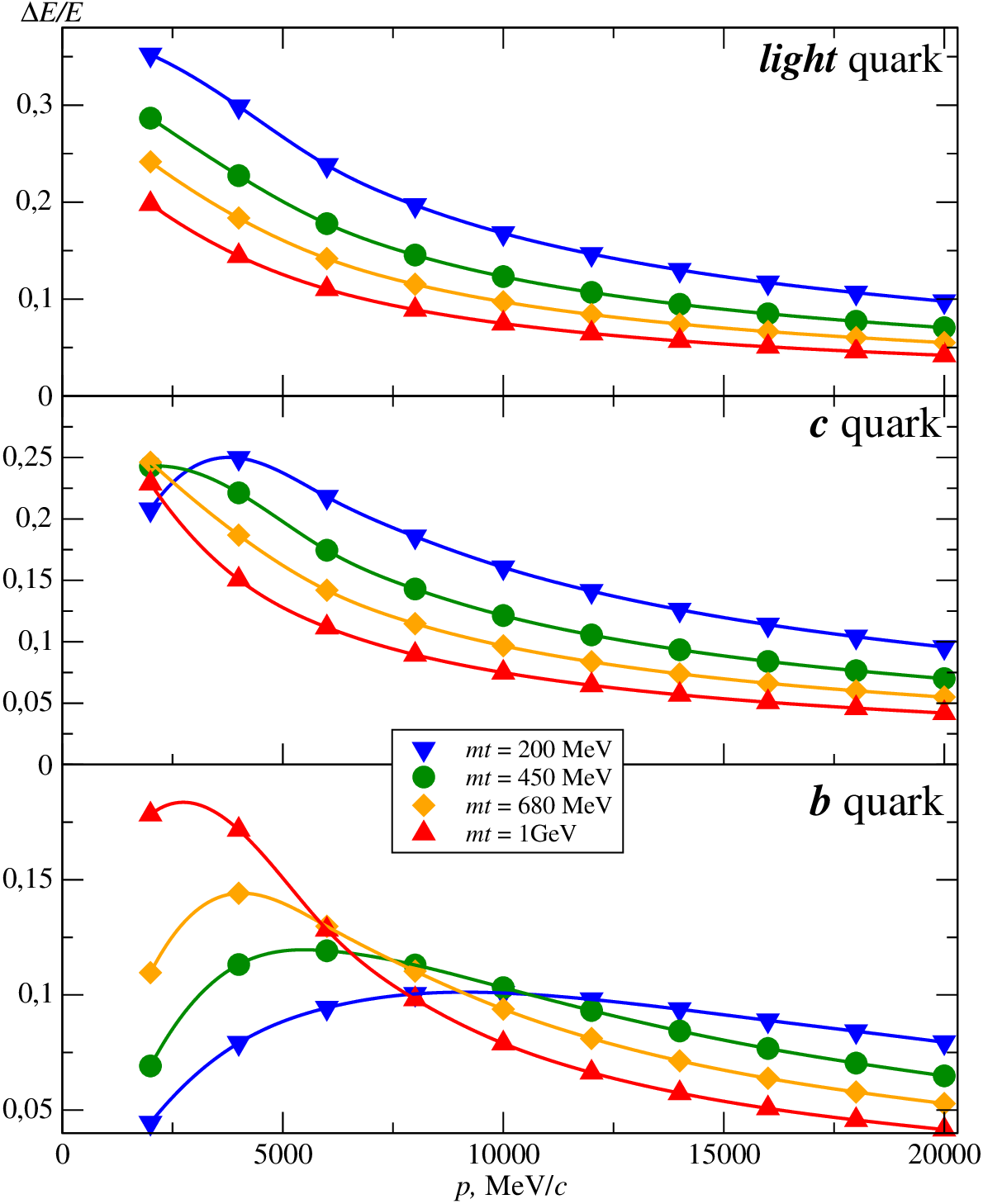}
\caption{Fractional energy loss as a function of projectile momentum. Symbols stand for calculated points, lines correspond to a cubic spline interpolation.} \label{dEtoE-fig}} \hfill
\parbox[t]{0.47\hsize}{\includegraphics[width=\hsize]{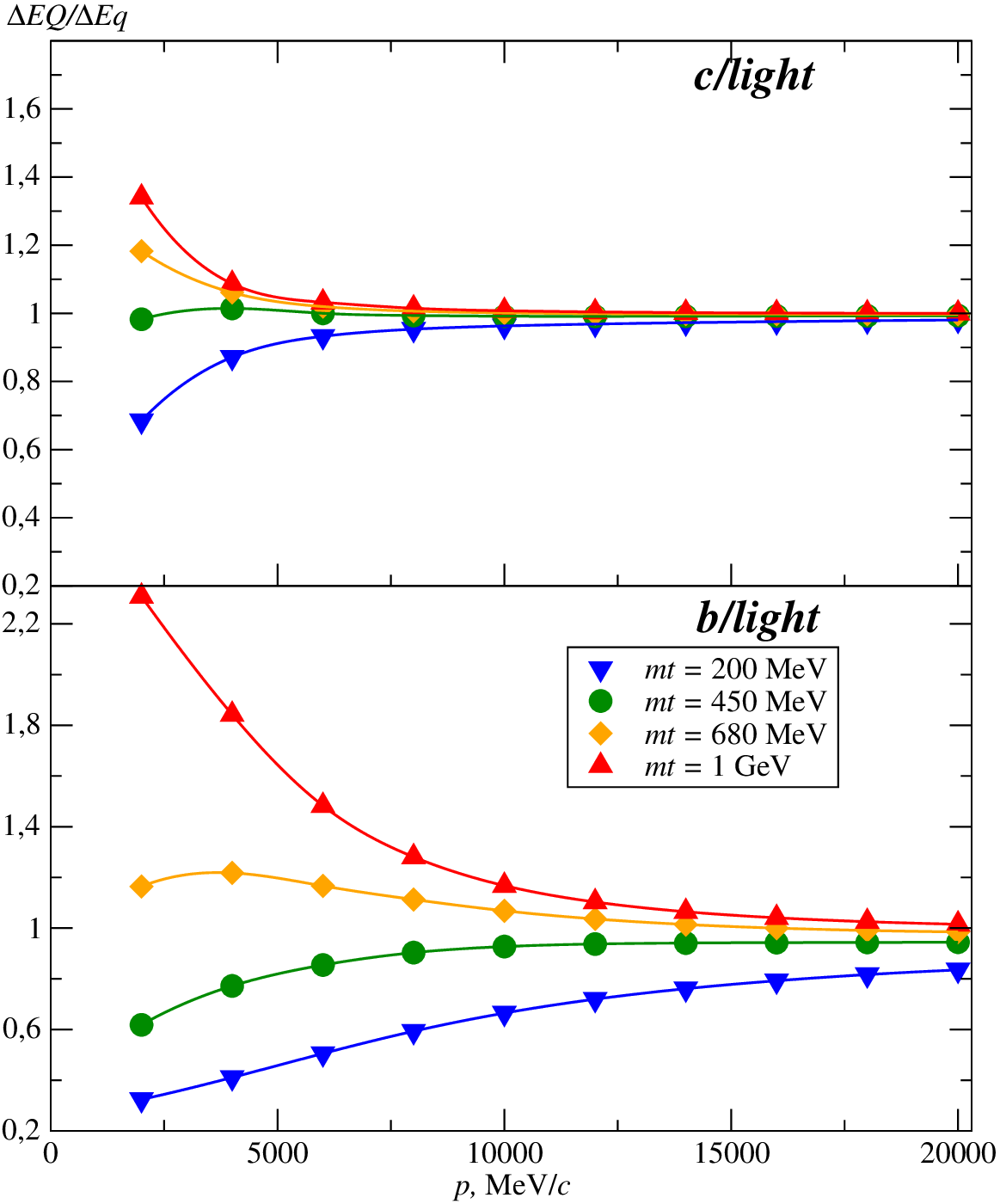} \caption{Ratio of heavy quark collisional energy loss to that of light quark. Symbols are calculated points, lines -- cubic spline interpolation.} \label{heavy-to-light}}
\end{figure}

As one can see from Fig.~\ref{dEtoE-fig}, the fractional collisional energy loss drops with increasing $m_t$ at high projectile mometum as expected. For $m_t=200$~MeV our results coincide with what has been obtained earlier for massless scatterers \cite{Djordjevic:2006tw}. What is interesting, is the less trivial dependence of heavy quark energy loss on $m_t$. Increase of $m_t$ leads to an increase in the energy loss for $b$ quarks, whereas energy loss for the light quarks drops. This effect s clearly illustrated at Fig.~\ref{heavy-to-light} we plot ratios of heavy quarks energy loss to that of light quark as a function of projectile momentum. At momentum $p=2\div 7$~GeV/$c$ one clearly observes how the mass hierarchy of collisional energy loss changes with $m_t$: at $m_t=1000$~MeV, it is opposite to that at $m_t=200$~MeV.

\begin{wrapfigure}[15]{o}{0.47\hsize}
\centering \vskip-3mm
\includegraphics[width=\hsize]{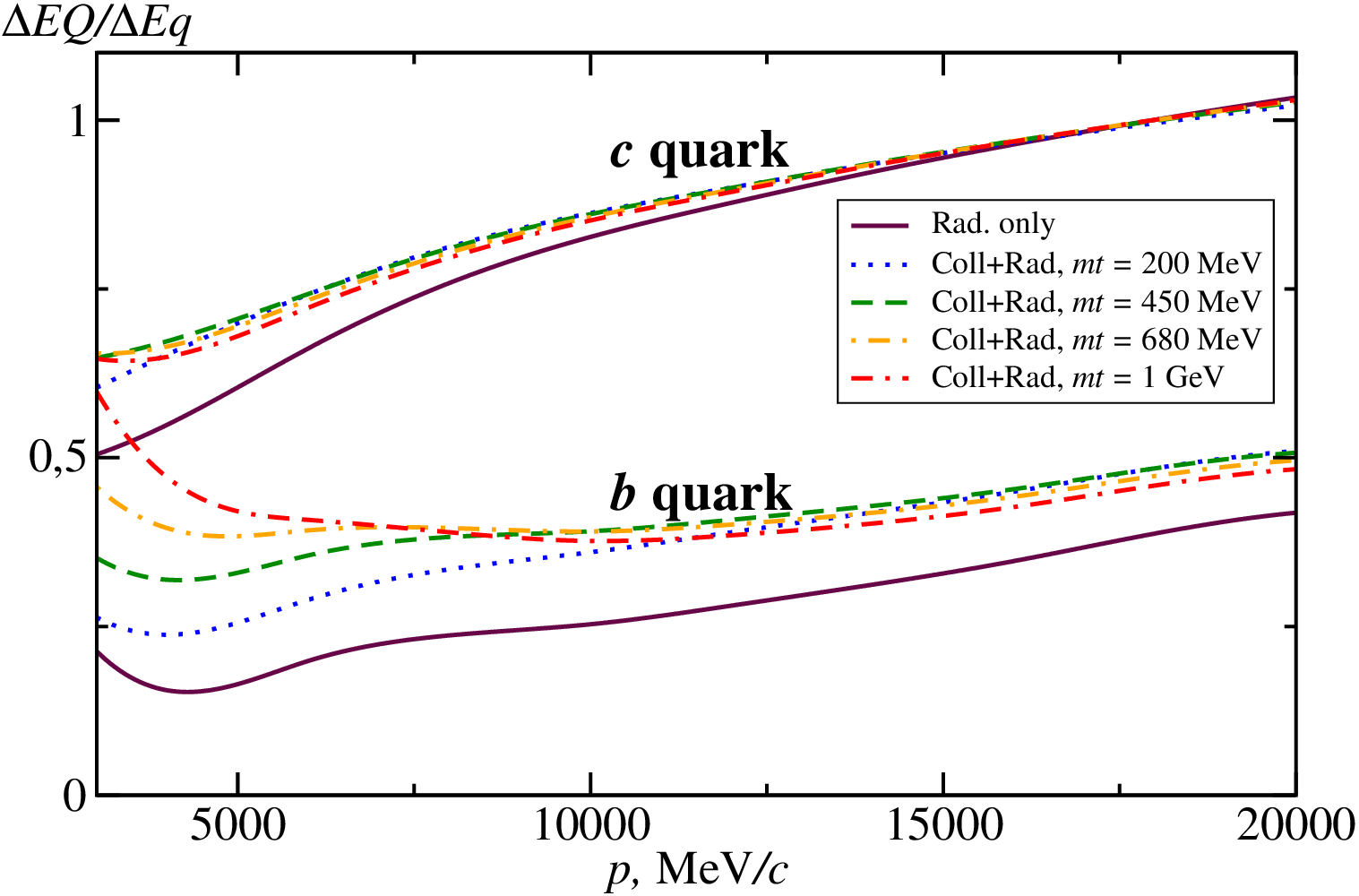}
\vskip-2mm
\caption{Ratios of the total (collisional plus radiative) energy loss of heavy quarks to that of light quark.} \label{tot-eloss}
\end{wrapfigure} 

The last question which we address here is the relative strength of collisional energy loss compared to the dominant radiative contribution, and how collisional energy loss could change mass hierarchy set up by the radiative. For this purpose we redraw Fig.~\ref{heavy-to-light} for the sum of radiative and collisional contributions. To evaluate radiative contribution we use ADSW quenching weights \cite{Armesto:2005iq} with $\hat q = 1$~GeV$^2$/fm and fix the pathlength at the same value of $L=5$~fm, the set of parameters being close to the one which produces the relevant factor 5 suppression in inclusive cross sections \cite{Dainese:2004te}.

As one observes at Fig.~\ref{tot-eloss}, if we include collisional contribution, total energy loss of heavy quarks relative to that of light quarks, increases with respect to the case of radiative energy loss only. This effect is especially pronounced for $b$ quarks the ratio rising from $\approx 0.1$ to $\approx 0.4$ for $m_t=1000$~MeV, so that mass ordering of radiative energy loss $\Delta E_Q \ll \Delta E_{\text{light }q}$ transforms into $\Delta E_Q \lesssim \Delta E_{\text{light }q}$ for $m_Q \gg m_{\text{light }q}$.


\section{Conclusion}

Within our simple model we have shown that elastic collisions of the fast projectile with partons of the medium can produce a significant contribution to the projectile energy loss. This collisional contribution was found to strongly depend on the ability of the medium to absorb the recoil which we effectively parametrized by some mass. Varying this parameter, it appeared to be possible not only to vary mass hierarchy of collisional energy loss alone, but also to reduce significantly the mass hierarchy set up by the radiative energy loss.

At the same time, applicability of the parameterization, as well as the relevant scale for the $m_t$ is still an open question.

\medskip
I thank Urs Wiedemann for his permanent interest and priceless contribution to the work.
This work was partially supported by INTAS 05-112-5031, grant RNP.2.2.2.2.1547, RBRF-CERN Nr~08-02-91004-CERN\_a and NFR Project 185664/V30.



\begin{thebibliography}{99}

\renewcommand{\itemsep}{0pt plus 2pt}

\bibitem{Djordjevic:2006tw}
  M.~Djordjevic,
  Phys.\ Rev.\  C {\bf 74} (2006) 064907
  [arXiv:nucl-th/0603066].

\bibitem{Wicks:2005gt}
  S.~Wicks, W.~Horowitz, M.~Djordjevic and M.~Gyulassy,
  Nucl.\ Phys.\  A {\bf 784} (2007) 426
  [arXiv:nucl-th/0512076].

\bibitem{Wicks:2007zz}
  S.~Wicks and M.~Gyulassy,
  J.\ Phys.\ G {\bf 34} (2007) S989
  [arXiv:nucl-th/0701088].


\bibitem{Kalashnikov:1979cy}
  O.~K.~Kalashnikov and V.~V.~Klimov,
  Sov.\ J.\ Nucl.\ Phys.\  {\bf 31}, 699 (1980)
  [Yad.\ Fiz.\  {\bf 31}, 1357 (1980)].

\bibitem{Klimov:1982bv}
  V.~V.~Klimov,
  Sov.\ Phys.\ JETP {\bf 55}, 199 (1982)
  [Zh.\ Eksp.\ Teor.\ Fiz.\  {\bf 82}, 336 (1982)].


\bibitem{Thoma:1990fm}
  M.~H.~Thoma and M.~Gyulassy,
  Nucl.\ Phys.\  B {\bf 351}, 491 (1991).

\bibitem{Braaten:1991we}
  E.~Braaten and M.~H.~Thoma,
  Phys.\ Rev.\  D {\bf 44}, 2625 (1991).


\bibitem{Dainese:2004te}
  A.~Dainese, C.~Loizides and G.~Paic,
  Eur.\ Phys.\ J.\  C {\bf 38} (2005) 461
  [arXiv:hep-ph/0406201].

\bibitem{Armesto:2005iq}
  N.~Armesto, A.~Dainese, C.~A.~Salgado and U.~A.~Wiedemann,  
  Phys.\ Rev.\  D {\bf 71} (2005) 054027
  [arXiv:hep-ph/0501225].  

\end{thebibliography}
\end{document}